\documentclass{WileyMSP-template}

\usepackage{bm}
\usepackage{tikz}
\usepackage{amsmath,amsthm,amssymb,amsfonts,amsbsy}
\usepackage{comment}
\usepackage{url}
\usepackage{xcolor}
\usepackage{caption}
\usepackage[normalem]{ulem}

\newcommand{\comm}[1]{}

\definecolor{red}{rgb}{0,0,0}

\begin{document}

\pagestyle{fancy}
\rhead{\includegraphics[width=2.5cm]{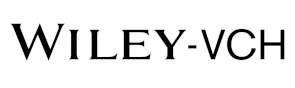}}

\title{Emergent morphogenesis via planar fabrication
	enabled by a reduced model of composites}

\maketitle

\author{Yupeng Zhang${^{\dag}}$}
\author{Adam Alon${^{\dag}}$}
\author{Mohammad Khalid Jawed*}

\dedication{}

\begin{affiliations}
Yupeng Zhang, Adam Alon, Mohammad Khalid Jawed\\
Department of Mechanical \& Aerospace Engineering\\
University of California Los Angeles\\
Los Angeles, CA 90095, USA\\
Email Address: khalidjm@seas.ucla.edu

\footnotetext[1]{$^\dag$ These authors contributed equally to this work.}

\end{affiliations}

\keywords{planar fabrication, bilayer composite, thermal actuation, reduced model, experimental validation}

\begin{abstract}

The ability to engineer complex three-dimensional shapes from planar sheets with precise, programmable control underpins emerging technologies in soft robotics, reconfigurable devices, and functional materials. Here, we present a reduced-order numerical and experimental framework for a bilayer system consisting of a stimuli-responsive thermoplastic sheet (Shrinky Dink) bonded to a kirigami-patterned, inert plastic layer. Upon uniform heating, the active layer contracts while the patterned layer constrains in-plane stretch but allows out-of-plane bending, yielding programmable 3D morphologies from simple planar precursors. Our approach enables efficient computational design and scalable manufacturing of 3D forms with a single-layer reduced model that captures the coupled mechanics of stretching and bending. Unlike traditional bilayer modeling, our framework collapses the multilayer composite into a single layer of nodes and elements, reducing the degrees of freedom and enabling simulation on a 2D geometry. This is achieved by introducing a novel energy formulation that captures the coupling between in-plane stretch mismatch and out-of-plane bending—extending beyond simple isotropic linear elastic models. Experimentally, we establish a fully planar, repeatable fabrication protocol using a stimuli-responsive thermoplastic and a laser-cut inert plastic layer. The programmed strain mismatch drives an array of 3D morphologies, such as bowls, canoes, and flower petals, all verified by both simulation and physical prototypes.
\end{abstract}

\section{Introduction}
Rapid and cost-effective manufacturing of 3D structures from planar 2D materials is challenging and the manufacturing methods are in significant demand in fields such as soft robotics, automobile manufacturing, biomedical devices, aerospace components, and deployable structures.
The existing structures fabricated from active and responsive materials have enabled applications in 
aerospace~\cite{furuya_concept_1992,hanaor_evaluation_2001,santiago-prowald_advances_2013}, medicine~\cite{du_plessis_dargentre_programmable_2018,banerjee_origami-layer-jamming_2020,babaee_temperature-responsive_2019}, and intelligent machines~\cite{wang_modular_2017,chi_bistable_2022},
but they remain primarily material-focused.
There is a growing need for scalable manufacturing frameworks that can accelerate real-world deployment. In this context, computational design -- particularly approaches that enable rapid simulation and optimization -- plays a pivotal role. Our work addresses this challenge by introducing a reduced computational model that enables efficient, accurate simulation of shape-morphing composites, leveraging algorithms originally developed in the computer graphics community~\cite{grinspun_discrete_2003}.

\medskip
A variety of pathways have been explored to 
{\color{red} 
achieve planar to three dimensional shape morphing, including 
soft kirigami composites for flexible deployable structures
\cite{zavodnik_soft_2024}, GAN-based inverse design \cite{brzin_using_2024,brzin_generative_2025},
shape optimization of electrostatic zipper actuator
\cite{akerson_mechanics_2023},
magnetic kirigami dome metasheet for dynamic shape shifting
\cite{chi_magnetic_2024},
multimodal shape evolution of shape memory polymers \cite{wu_programmable_2025},
}
tailored material designs~\cite{dudek_shape_2025}, composite structures~\cite{daynes_review_2013,rodriguez_shape-morphing_2016}, microstructurally engineered polymers~\cite{aharoni_universal_2018, jin_shape-morphing_2022, studart_bioinspired_2014}, origami-inspired mechanisms~\cite{hawkes_programmable_2010, tolley_self-folding_2014}, and architected metastructures~\cite{boston_spanwise_2022, risso_highly_2022, zhang_shape-morphing_2025}. 
{\color{red} 
In particular, Brzin and Brojan \cite{brzin_using_2024}
developed a mechanical analog framework for the inverse design of soft morphing composite beams based on strain mismatch induced by temperature actuation.
}
The transformation from 2D to 3D forms is governed by the interplay between stretching and bending~\cite{armon_geometry_2011}, where strain mismatches embedded in planar preforms can lead to a variety of 3D shapes~\cite{aharoni_universal_2018, ware_voxelated_2015, giudici_multiple_2022, modes_gaussian_2011}. In particular, bilayer composites comprising a stimuli-responsive material bonded to a patterned kirigami layer~\cite{mungekar_directed_2025} offer a scalable, 
manufacturing driven strategy for programmable shape morphing across a broad range of applications, such as deployable structures, patient-specific biomedical devices, and adaptive surfaces. In this work, we use a bilayer system composed of a Shrinky Dink thermoplastic and an inert kirigami-patterned plastic as a model platform for programmable morphable surfaces. Our reduced-order numerical method, demonstrated on this system, can be readily augmented to simulate and guide the design of other classes of shape-morphing composite shells.

\begin{figure}[!htb]
	\begin{center}
		{\includegraphics[width=6.5in]{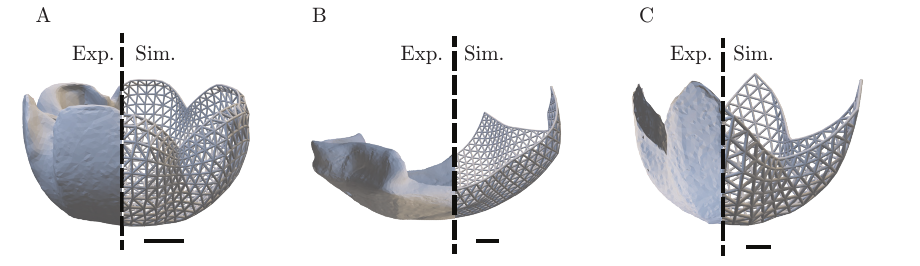}}
		\caption{
			Examples demonstrating the validation between 3D shapes manufactured by physical experiments using the protocol developed here and the ones generated by numerical simulations using the proposed energy in Equation~\ref{eq:bilayer_energy}. Scalebar: 1 cm.}
		\label{fig:1}
	\end{center}
\end{figure} 

\medskip
In this paper, we introduce a numerical framework for efficient and accurate prediction of stimulus-driven 3D shape morphing in bilayer composite shells. Unlike conventional approaches that require explicit modeling of multiple layers, we represent the entire layered composite using a single layer of nodes and formulate an effective energy model that captures the coupled mechanics of the bilayer system. Our method builds on the discrete shells approach~\cite{grinspun_discrete_2003}, where the total elastic energy of a thin structure is computed from in-plane stretching (changes in edge lengths) and out-of-plane bending (dihedral angles between adjacent triangular elements), each referenced to rest lengths and rest angles that define the undeformed, stress-free configuration. We advance this framework by allowing these rest lengths and rest angles to evolve explicitly as functions of the applied stimulus (heat in this work), enabling the model to capture permanent, stimulus-driven deformations such as those caused by thermal contraction in heat-shrinking layers. 
This energy-based approach enables accurate predictions of complex 3D configurations arising from 2D patterned precursors, without resorting to computationally expensive full 3D simulations.
{\color{red}
A direct comparison of computational cost between the discrete elastic shell (DES) and finite element (FE) methods is presented at the end of Section~\ref{sec:FEA}.
}
We validate our framework by comparing numerical predictions to physical experiments on bilayer composites composed of a heat-shrinking thermoplastic (Shrinky Dink) and a layer of 3D printed PLA (polylactic acid).

\medskip
\textbf{Figure~\ref{fig:1}} shows representative 3D shapes fabricated in physical experiments (left) and their numerically simulated counterparts (right). The experimental fabrication protocol builds directly on prior work~\cite{mungekar_directed_2025}, while the present study provides, for the first time, a predictive numerical framework for these systems. Our main contributions are:

\begin{itemize}
    \item Introduction of a reduced-order modeling approach in which the composite is represented by a single layer of nodes; this framework is broadly applicable to a variety of layered or architected material systems, beyond bilayer composites.
    \item Explicit incorporation of external stimuli by formulating the rest lengths and rest angles (i.e., the stress-free, undeformed configuration) as functions of the applied stimulus, enabling accurate modeling of permanent, stimulus-driven morphing.
    \item Quantitative demonstration that the numerical predictions closely match experimental measurements across a range of complex shapes and kirigami patterns; see Figure~\ref{fig:1}.
\end{itemize}

\medskip
We begin by detailing the experimental approach, including the materials, methods, and fabrication protocols.
Next, we introduce our new energy formulation and describe its numerical implementation within the discrete shells algorithm. 
To demonstrate the effectiveness of our approach, we study three representative kirigami patterns and analyze their resulting 3D shapes through both numerical simulations and physical experiments. We quantify the differences between the simulated and experimental shapes at various stages of the stimulus, and find excellent agreement, thereby validating the predictive capability of our framework.

\section{Results and Discussion}

\subsection{Emergent 3D Shapes via Planar Fabrication of Bilayer Composites}

The experimental approach exploits the rapid contraction of thermally responsive polystyrene (Shrinky Dink) together with the geometric programmability enabled by 3D printing. Polystyrene is selected for its reliable shrinkage (up to 60\% reduction in length), low cost, and safe processing in standard laboratory ovens. The shrinkage ratio can be precisely tuned by adjusting the heating time, providing an additional degree of control over the final shape. For programmable morphing, we digitally design and directly 3D print the inert plastic (PLA) layer in the desired kirigami pattern -- a network of openings and solid regions that enables localized bending and selectively constrains shrinkage. This additive manufacturing approach ensures high precision and reproducibility, which are difficult to achieve with manual cutting methods such as scissors. The patterned PLA is then bonded to the polystyrene sheet, allowing for spatial control of deformation and a customizable mechanical response.

\begin{figure}[!htb]
\begin{center}
{\includegraphics[width=6.5in]{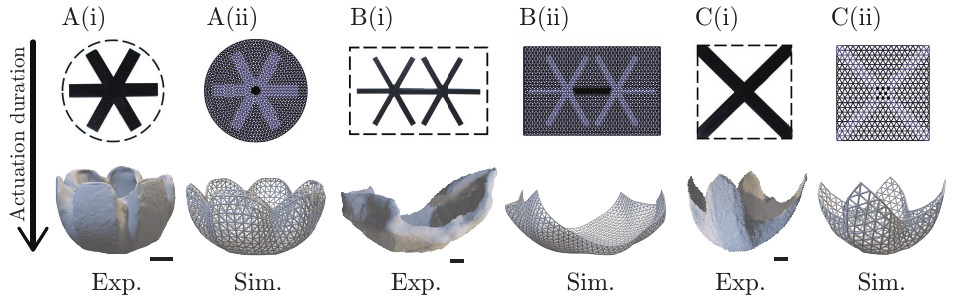}}
\caption{
Examples demonstrating the morphing of 2D bilayer composites into 3D shapes via thermal actuation.
The first row shows the 2D designs: each design consists of a specific kirigami pattern (PLA layer) on a thermoplastic (Shrinky Dink) substrate of defined shape.
The second row displays the resulting 3D shapes.
Subfigures A, B, and C correspond to three representative cases: bowl, canoe, and flower petal morphologies, respectively.
For each design, the left column (i) presents results from physical experiments, and the right column (ii) shows the corresponding numerical simulations using the new energy model in Equation~\ref{eq:bilayer_energy}.
}
\label{fig:2}
\end{center}
\end{figure}

\medskip
\textbf{Figure~\ref{fig:2}} presents three representative examples demonstrating the morphing of 3D shapes from planar bilayer composites via thermal actuation. In the first row, panels A(i)–C(i) show physical scans of the digitally designed inert kirigami layer (solid black regions) alongside the shapes of the Shrinky Dink substrate (dashed boundaries) for each 2D precursor, 
while panels A(ii)-C(ii) display the corresponding numerical meshes.
The second row shows the resulting physical and numerically simulated 3D shapes. Depending on the kirigami pattern, the flat 2D precursor undergoes stimulus-driven transformation in the oven to form a variety of 3D shapes, such as a bowl, a canoe, and flower petals. The close agreement between the experimentally scanned shapes and the simulation results demonstrates the predictive capability of our numerical framework.

\medskip
To fabricate the morphable bilayer composites, we first design a kirigami pattern for the PLA layer based on heuristics that relate the pattern geometry to the intended 3D morphed shape. The chosen pattern is then digitally modeled and 3D printed as a thin PLA structure. This PLA layer is bonded to the polystyrene (Shrinky Dink) sheet using a two-part Gorilla epoxy. The high bending and stretching stiffness of PLA constrains shrinkage in the bonded regions, allowing the patterned geometry to locally direct where in-plane contraction is allowed or resisted. As a result, the composite exhibits spatially varying stiffness, which can be further tuned by adjusting the thickness of the PLA layer to achieve complex 3D surfaces. 

\begin{figure}[!htb]
	\begin{center}
		{\includegraphics[width=6.5in]{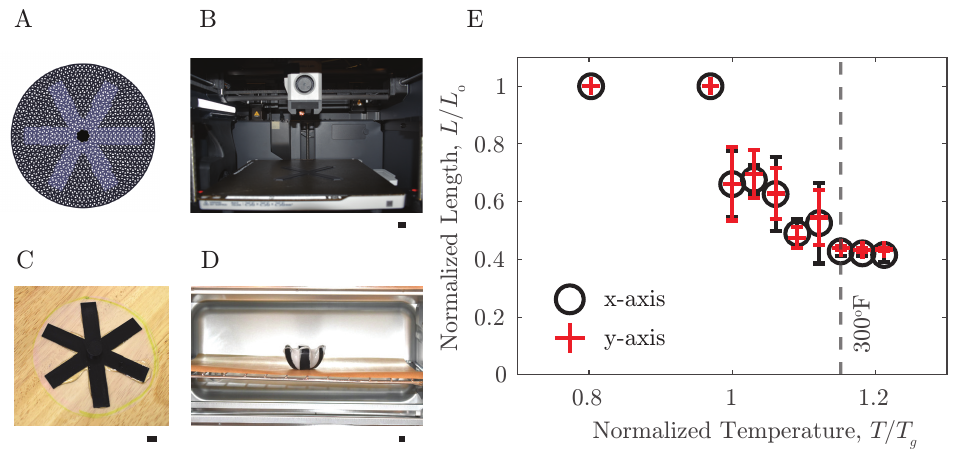}}
\caption{
Experimental setup. 
(A) Design of the kirigami pattern (inert PLA layer) and selection of the substrate domain (thermoplastic Shrinky Dink sheet). 
(B) 3D printing of the kirigami pattern. 
(C) Assembly of the bilayer composite by bonding the 3D-printed PLA pattern to the substrate. 
(D) Thermal actuation in an oven for shape morphing. 
(E) Measured variation of the shrink ratio ($L/L_0$) as a function of normalized temperature ($T/T_g$), where $T_g = 200^\circ$F ($366.5$K) is the glass transition temperature of polystyrene. The sheet begins to shrink when $T/T_g > 1$, and the shrink ratio stabilizes above $T/T_g \approx 1.15$ (indicated by the vertical dashed line). Scale bars in (B), (C), and (D) denote 1 cm.
}
		\label{fig:3}
	\end{center}
\end{figure}

\medskip
\textbf{Figure~\ref{fig:3}} illustrates the complete fabrication process. First, the kirigami pattern is designed based on the desired final shape; for example, Figure~\ref{fig:3}A shows a six-arm star on a circular thermoplastic sheet. The pattern is then digitally modeled with the prescribed thickness using Fusion \cite{autodesk_inc_fusion_2025} and fabricated by 3D printing, as shown in Figure~\ref{fig:3}B. Next, the printed PLA pattern is bonded to a precisely cut thermoplastic substrate, and the assembled bilayer composite is cured for over 10 hours to ensure robust adhesion (Figure~\ref{fig:3}C).
After curing, thermal actuation is used to induce shape morphing, as shown in Figure~\ref{fig:3}D, by heating the planar composite in an oven for approximately 15 seconds. The response of the Shrinky Dink layer is characterized by measuring its shrinkage as a function of normalized temperature (Figure~\ref{fig:3}E), where the temperature is scaled by the glass transition temperature ($T_g$) of polystyrene, an amorphous thermoplastic without a sharp melting point. The sheet begins to shrink when $T/T_g > 1$ (temperatures are measured in Kelvins), but the shrink ratio $L/L_0$ does not stabilize until it exceeds 1.15, corresponding to temperatures above $300^\circ$F, as indicated by a vertical dashed line in Figure~\ref{fig:3}E. Consequently, all experiments are performed at temperatures greater than $300^\circ$F. 

\medskip
Using this protocol, we explored a range of additively manufactured patterns with varying dimensions and sheet shapes. This paper highlights three representative designs (Figure~\ref{fig:1}A–C), each defined by a distinct combination of substrate domain and kirigami pattern: (A) a six-arm star kirigami pattern on a circular Shrinky Dink substrate, (B) a two-star kirigami pattern on a rectangular substrate, and (C) a cross kirigami pattern on a square substrate. After thermal activation at $300^\circ$F for approximately 15 seconds, the planar composites reliably morph into 3D shapes, which are then reconstructed and analyzed through 3D scanning. In the following section, we present our numerical method for predicting these experimentally observed 3D shapes.

\subsection{Numerical Simulation with a Reduced Model}
\label{sec:numerics}
Motivated by the experimental observations of programmable shape morphing in bilayer composites -- comprising a 3D printed PLA kirigami layer bonded to a Shrinky Dink sheet -- we propose a new energy formulation that couples planar stretching with out-of-plane deformation. We first briefly review the discrete elastic shells method in Section~\ref{sec:DEP}, and then introduce our reduced energy model for composite shells in Section~\ref{sec:energy}.

\subsubsection{Discrete Elastic Shells Method}
\label{sec:DEP}

\begin{figure}[!htb]
	\begin{center}
		{\includegraphics[width=6.5in]{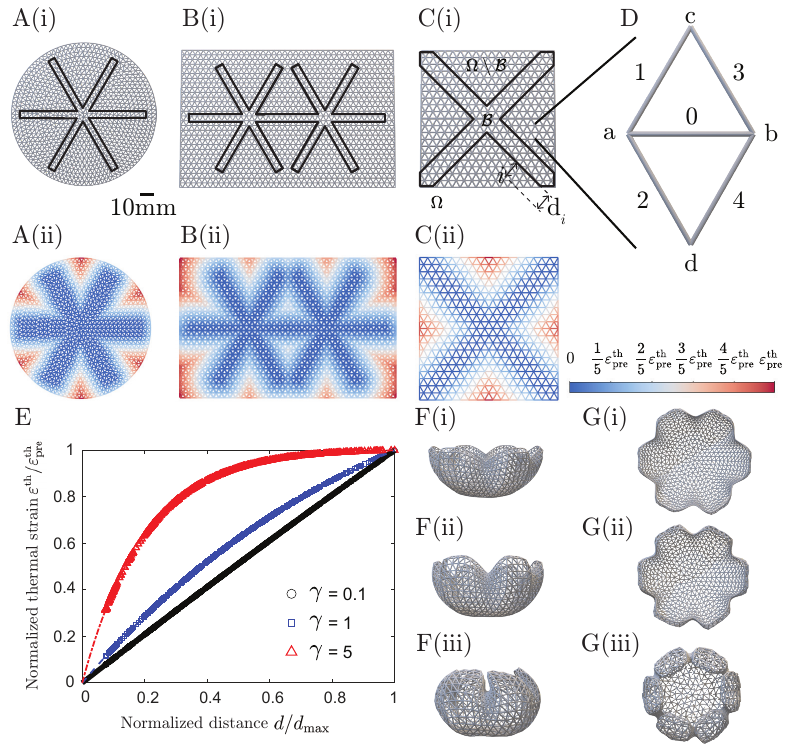}}
		\caption{Reference-configurations meshes for designs A-C in Figures~\ref{fig:1} and \ref{fig:2}.
        A-C(i) Thermoplastic PLA bilayer boundaries (solid lines) on discrete mesh.
        A-C(ii) The distribution of thermal strains as defined in Equation~\ref{eq:thermal_strain}. See Section~\ref{sec:energy} for definitions of
        $\Omega$, $\mathcal{B}$, $\varepsilon_{\rm pre}^{\mathrm{th}}$, and $d_i$.
        D. Local node and edge IDs for a bending pair.
        The number of nodes $N_n$, number of edges $N_e$, and number of triangular elements $N_t$, $(N_n, N_e, N_t)$ are (970, 2800, 1831) for pattern A, (1215, 3505, 2291) for pattern B, and (388, 1087, 700) for pattern C, respectively.
        {\color{red} E. Variation of normalized thermal strain $\varepsilon^{\mathrm{\rm th}}/\varepsilon_{\rm pre}^{\mathrm{\rm th}}$ with normalized distance ${d}/{d_{\max}}$ for different values of $\gamma$, corresponding to the characteristic length $L_c=d_{\max}/\gamma$ in Equation~\ref{eq:Lc}. 
        F(i)-(iii) Deformed configurations of pattern A with $\varepsilon_{\rm pre}^{\mathrm{\rm th}}=0.7$ for $\gamma=0.1$, 1, and $5$, respectively. G. Top views of the corresponding configurations shown in F.}
        }
		\label{fig:4}
	\end{center}
\end{figure}

The discrete elastic shells framework provides a discrete differential geometry-based approach for modeling thin shells, where the thickness is much smaller than the in-plane dimensions. In this formulation, the structure is represented as a triangular mesh with $N$ nodes connected by $N_e$ edges, as shown in \textbf{Figure~\ref{fig:4}}. The specific meshes used to simulate the three representative examples in this work are shown in Figures~\ref{fig:4}A–C. Each node (e.g., $a, b, c, d$ in Figure~\ref{fig:4}D) has three spatial degrees of freedom, corresponding to its position in 3D space, and the full configuration is described by the vector $\mathbf X$ of size $3N$. An edge is the connection between two nodes, labeled as $0$, $1$, $2$, $3$, and $4$ in Figure~\ref{fig:4}D. Edges that are shared by two triangles are referred to as hinges (for example, edge $0$ is the common edge of triangles $abc$ and $adb$). The mesh contains $N_h$ such hinges. At every hinge, the angle between the normals of the adjacent triangles is defined as the dihedral angle. Edges can undergo stretching, and hinges can undergo bending; these deformations store stretching and bending energies, respectively. The corresponding elastic energy expressions are introduced later in this section.

\medskip
The equations of motion for the DOF vector $\mathbf X$ are given by
\begin{equation}
    \mathbb{M} \ddot{\mathbf X} + \nabla E = \mathbf F_\textrm{ext},
    \label{eq:EOM_continuous}
\end{equation}
where $\mathbb{M}$ is the (diagonal) mass matrix, $\dot{\mathbf X}$ and $\ddot{\mathbf X}$ denote the first and second time derivatives of $\mathbf X$, $\nabla E$ is the gradient of the total elastic energy with respect to the DOF vector (i.e., the negative of the elastic force), and $\mathbf F_\textrm{ext}$ is the external force vector. To advance the simulation from $t = t_k$ to $t_{k+1} = t_k + \Delta t$, we employ the implicit Euler method and discretize Equation~\ref{eq:EOM_continuous} as
\begin{equation}
    \mathbb{M} \frac{1}{\Delta t} \left( \frac{\mathbf X (t_{k+1}) - \mathbf X (t_k)}{\Delta t} - \dot{\mathbf X} (t_k) \right) + \nabla E(\mathbf X (t_{k+1})) = \mathbf F_\textrm{ext}(t_{k+1}).
    \label{eq:EOM_discrete}
\end{equation}
Given the positions $\mathbf X (t_k)$ and velocities $\dot{\mathbf X} (t_k)$ at the previous time step, the new positions $\mathbf X (t_{k+1})$ are computed by solving this nonlinear system via the Newton–Raphson method. The Jacobian matrix for each Newton iteration is
\begin{equation}
    \mathbf{J} = \frac{\mathbb{M}}{\Delta t^2} + \nabla^2 E(\mathbf X) - \nabla \mathbf F_\textrm{ext},
    \label{eq:jac}
\end{equation}
where $\nabla^2 E$ is the Hessian of the elastic energy with respect to nodal positions. Since our study focuses on the static equilibrium shape, the inertial term $\mathbb{M} \ddot{\mathbf X}$ is neglected and the external force $\mathbf F_\textrm{ext}(t_{k+1})$ is set to zero. In this case, the final configuration corresponds to a minimum of the total elastic energy. The velocity at $t_{k+1}$ can be calculated simply from $\dot{\mathbf X} (t_{k+1}) = \left( \mathbf X (t_{k+1}) - \mathbf X (t_k) \right) / \Delta t$.

\medskip
In the discrete shells framework, the total elastic energy is the sum of stretching energies (associated with each edge) and bending energies (associated with each hinge)~\cite{savin_growth_2011}:
\begin{equation}
    E\bigl(\bm X\bigr)
    =\sum_{i=1}^{N_e}\frac12\,k_s\,\varepsilon_i^2
    +
    \sum_{j=1}^{N_h}\frac12\,k_b\, \left( \theta_j - \bar{\theta}_j \right)^2,
    \label{eq:energyElastic}
\end{equation}
where $\varepsilon_i$ (axial strain for edge $i$) and $\theta_j$ (dihedral angle at hinge $j$) are defined below as macroscopic, geometrically invariant strain measures, and $\bar{\theta}_j$ denotes the rest dihedral angle at hinge $j$, i.e., the dihedral angle in the reference (undeformed) configuration. For a flat plate, $\bar{\theta}_j = 0$, while for a curved shell, $\bar{\theta}_j$ may be nonzero and spatially varying.

\medskip
The strains are fully determined by the DOF vector $\mathbf X$ and are invariant with respect to the global coordinate system. Specifically, the axial strain for each edge is $\varepsilon_i = l_i/l_{0,i} - 1$, where $l_i$ is the current edge length and $l_{0,i}$ is its rest (reference) length in the undeformed mesh. The dihedral angle $\theta_j$ is defined as the angle between the normals of the two triangles sharing hinge $j$. The stretching stiffness $k_s$ and bending stiffness $k_b$ are
\begin{equation}
  \begin{aligned}
    k_s &= \frac{\sqrt{3}}{2}\,Y\,h\,l_0^2, \\
    k_b &= \frac{2}{\sqrt{3}}\,D, \quad D=\frac{Y\,h^3}{12}\,.
  \end{aligned}
  \label{eq:ks_kb}
\end{equation}
where $Y$ is the effective Young's modulus, $h$ is the shell thickness, and $l_0$ is the mean edge length. 
These discrete expressions are constructed such that, in the limit of mesh refinement, the model recovers the continuum F\"{o}ppl--von K\'{a}rm\'{a}n equations for nonlinear thin plates. The three composites have domains $\Omega$ of
100 mm radius (pattern A), 100 mm $\times$ 156 mm (pattern B),
and 100 mm $\times$ 100 mm (pattern C), Figures~\ref{fig:2} and \ref{fig:4}.
With the mesh configurations in Figure~\ref{fig:4}, we have
$l_0=3.2$ mm, $4.0$ mm, and $5.8$ mm for A, B, and C, respectively. 
The thicknesses of two layers and their corresponding effective stretching and bending stiffness values, $k_s$ and $k_b$, are given in Section~\ref{sec:validation}.

\medskip
In our model, the computational mesh consists of two distinct types of regions, Figure~\ref{fig:4} A-C(i):
(i) bilayer regions $\mathcal{B}$, where both the Shrinky Dink and the inert kirigami layer are present, and (ii) single-layer regions $\Omega\setminus\mathcal B$, where only the Shrinky Dink substrate remains (corresponding to the cut-out areas of the kirigami pattern). The material properties -- such as Young's modulus $Y$, and consequently the stretching and bending stiffnesses $k_s$ and $k_b$ -- are assigned locally to each mesh element based on its region: bilayer regions use effective properties of the combined layers, while single-layer regions use properties of the Shrinky Dink alone. The specific numerical values used for all material parameters are provided in the Section~\ref{sec:validation}.

\medskip
While the assignment of material properties enables accurate representation of spatial heterogeneity in the composite, classical discrete plate models are limited to flat, stress-free reference states. In contrast, discrete shells extend this framework by permitting nonzero natural (rest) curvature, achieved by introducing nonzero rest dihedral angles. This is implemented by replacing the bending term $\frac12\,k_b\, \theta_j^2$ with $\frac12\,k_b\, \left( \theta_j - \bar{\theta}_j \right)^2$ in the energy expression. Such an extension is essential for capturing the permanent, stimulus-induced shape changes observed in our bilayer composites. Building on these principles, we next present our reduced energy model for programmable composite shells, which incorporates stimulus-dependent rest configurations to predict the full range of morphable shapes observed in experiments.

\subsubsection{Energy Formulation for Stimulus-Responsive Composite Shells}
\label{sec:energy}

The core innovation of our approach is to extend the discrete shells framework so that the elastic energy depends explicitly on both the DOF vector $\mathbf X$ -- representing the shape of the structure -- and the applied external stimulus, such as temperature. This enables us to efficiently model the permanent, stimulus-driven morphing of bilayer shells using only a single layer of mesh nodes, without the need for computationally intensive 3D elements. The key physical mechanism is the non-uniform thermal strain induced in the bilayer composite: when a 3D-printed PLA pattern is bonded to a Shrinky Dink substrate, only the Shrinky Dink contracts under heating, while the PLA remains largely inert. This mismatch in planar contraction leads to locally varying natural curvatures -- captured in our model as stimulus-dependent rest dihedral angles at the mesh hinges. By properly designing the kirigami pattern, we control which regions bend and by how much, enabling programmable shape transformation. Building on the framework established above, we generalize the energy expression in Equation~\ref{eq:energyElastic} by allowing the rest dihedral angle $\bar{\theta}_j$ to depend explicitly on the applied thermal stimulus, $T$.

\medskip
In our reduced model, the total elastic energy is written as
\begin{equation}
    E(\bm X, T) = \sum_{i=1}^{N_e} \frac{1}{2} k_s \left[\varepsilon_i (\bm X) - \varepsilon_i^{\mathrm{th}} (T) \right]^2
    + \sum_{j=1}^{N_h} \frac{1}{2} k_b \left[ \theta_j (\bm X) - \beta l_0 \Delta\varepsilon_j (\bm X, T) \right]^2,
    \label{eq:bilayer_energy}
\end{equation}
where $\varepsilon_i^{\mathrm{th}}(T)$ is the stimulus-induced thermal strain in the Shrinky Dink layer (measured from experiments, cf. Figure~\ref{fig:3}E). The first term accounts for in-plane stretching, while the second term models out-of-plane bending. Here, $\theta_j(\bm X)$ is the dihedral angle at hinge $j$, and $\beta l_0 \Delta\varepsilon_j(\bm X, T)$ replaces the rest dihedral angle $\bar{\theta}_j$ in Equation~\ref{eq:energyElastic}, quantifying the local mismatch in planar strain across the two triangles sharing each hinge. 
The explicit expression for $\Delta \varepsilon_j$ is provided later in this section. We write $\varepsilon_i(\bm X)$, $\theta_j(\bm X)$, and $\Delta \varepsilon_j(\bm X, T)$ to emphasize that these quantities are computed from the DOF vector $\bm X$ and, where relevant, the applied stimulus $T$. Similarly, $\varepsilon_i^{\mathrm{th}}(T)$ denotes the thermal strain as a function of $T$, computed directly from the experimentally measured contraction of the Shrinky Dink sheet at each temperature (see Figure~\ref{fig:3}E).

\medskip
{\color{red}
The parameter $\beta$ quantifies the strength of coupling between in-plane strain mismatch and out-of-plane bending and has units of inverse length.
We consider a bilayer slender beam with traction-free boundary conditions subjected to uniform tensile strains $\varepsilon_1$ and $\varepsilon_2$ in layer 1 and layer 2, respectively.
When the strain mismatch $\Delta \varepsilon=\varepsilon_1-\varepsilon_2$ is zero, the beam remains flat, i.e. corresponds to a zero natural curvature.
Whereas a nonzero mismatch $\Delta \varepsilon \neq 0$ induces a nonzero natural curvature $\bar{\kappa} \ne 0$.
In the discrete shell formulation, the curvature $\bar{\kappa}$ is related to the bending angle $\bar{\theta}$ between adjacent elements by \cite{grinspun_discrete_2003}
\begin{equation}
\bar{\kappa} = \bar{\theta}/l_0,
\label{eq:curvature1}
\end{equation}
where $l_0$ is the element length.
From the energy in Equation~\ref{eq:bilayer_energy}, we have $\bar{\theta}=\beta l_0 \Delta \varepsilon$. Substitute this into Equation~\ref{eq:curvature1}, we obtain
\begin{equation}
\bar{\kappa}=\beta \Delta \varepsilon,
\label{eq:curvature2}
\end{equation}
indicating that $\beta$ directly relates local strain mismatch to local curvature per unit length.

\medskip
In principle, $\beta$ could be calibrated experimentally by measuring curvature as a function of strain mismatch in a controlled bilayer beam experiment. However, under the non-uniform thermal actuation considered here, a spatially resolved experimental calibration of $\beta$ is not straightforward and is beyond the scope of the present study. Therefore, in this work we adopt the choice $\beta=1/l_0$, such that the dimensionless quantity $\beta l_0=1$.
This choice minimizes sensitivity to mesh discretization while preserving the correct scaling between strain mismatch and curvature. A more detailed experimental calibration of spatially varying $\beta$ fields will be pursued in future work.

\subsubsection{Thermal Strain and Temperature Change Distribution}

\medskip
The thermal strain arises from shrinkage induced by temperature change and is described by the classical relation \cite{timoshenko_goodier_1970}
\begin{equation}
\varepsilon_{\rm pre}^{\rm{th}} = \alpha \Delta T.
\end{equation}
Experimental observations indicate that deformation near the free boundary of the Shrinky Dink sheet is more pronounced than in regions constrained by the PLA kirigami bilayer. 
This suggests a spatially nonuniform thermal strain field caused by partial constraint of shrinkage near the bilayer region.
To capture this effect, we quantify the distance from an edge midpoint $i$ to the closest Shrinky Dink-PLA bilayer boundary as $d_i$, and define $d_{\max}$ as the maximum such distance over the domain,
\begin{equation}
    d_i = \min_{j\in\mathcal{B}}\bigl\lVert \mathbf{x}_i - \mathbf{x}_j\bigr\rVert,
    \quad
    d_{\max} = \max_{i\in \Omega\setminus \mathcal{B}}d_i,
    \label{eq:define_d}
\end{equation}
where $\Omega$ denotes the set of all edge midpoints in the composite domain,
and $\mathcal{B} \subset \Omega$ denotes the subset lying within the Shrinky Dink-PLA bilayer region.
Here, $\mathbf{x}_i$ is the midpoint coordinate of edge $i\in\Omega\setminus\mathcal B$, 
and $\mathbf{x}_j$ that of edge $j\in\mathcal B$.

\medskip
To prescribe a thermal strain field consistent with experimental observations, we require that it satisfies the following conditions:\\
(i) $\varepsilon^{\mathrm{th}}(0)=0$ at the kirigami bilayer boundary.\\
(ii) $\varepsilon^{\mathrm{th}}(d_{\max})=\varepsilon_{\mathrm{pre}}^{\mathrm{th}}$, corresponding to free shrinkage far from the bilayer boundary.\\
(iii) $\varepsilon^{\mathrm{th}}(d)$ increases monotonically with distance $d$.

\medskip
A general form satisfying these requirements is an exponential profile
\begin{equation}
	\varepsilon^{\mathrm{th}}(d)
	=
	\varepsilon_{\mathrm{pre}}^{\mathrm{th}}
	\,
	\frac{1 - \exp\!\left(-{d}/{L_c}\right)}
	{1 - \exp\!\left(-{d_{\max}}/{L_c}\right)},
    \quad d \in [0,d_{\max}],
	\label{eq:thermal_exp}
\end{equation}
where $L_c>0$ is a characteristic decay length controlling the spatial localization of thermal strain,
and $\varepsilon_{\mathrm{pre}}^{\mathrm{th}}$ is the prescribed thermal strain parameter.

\medskip
We introduce a dimensionless localization parameter $\gamma \in (0, \infty)$ defined by
\begin{equation}
	L_c = {d_{\max}}/{\gamma}.
    \label{eq:Lc}
\end{equation}
With this, Equation~\ref{eq:thermal_exp} can be expressed in nondimensional form as
\begin{equation}
	\varepsilon^{\mathrm{th}}(d/d_{\max})
	=
	\varepsilon_{\mathrm{pre}}^{\mathrm{th}}
	\,
	\frac{1 - \exp(-\gamma d/d_{\max})}
	{1 - \exp(-\gamma)}.
	\label{eq:thermal_dimensionless}
\end{equation}

\medskip
As shown by the asymptotic analysis in Appendix~\ref{app:limit}, this formulation admits two limiting cases.
In the limit $\gamma \to 0$,
Equation~\ref{eq:thermal_dimensionless} reduces to
\begin{equation}
	\lim_{\gamma \to 0}
	\varepsilon^{\mathrm{th}}(\frac{d}{d_{\max}})
	=
	\varepsilon_{\mathrm{pre}}^{\mathrm{th}}
	\frac{d}{d_{\max}},
	\label{eq:thermal_linear}
\end{equation}
which corresponds to a linear thermal strain relation.
In contrast, in the limit $\gamma \to \infty$,
Equation~\ref{eq:thermal_dimensionless} reduces to
\begin{equation}
	\lim_{\gamma \to \infty}
	\varepsilon^{\mathrm{th}}(\frac{d}{d_{\max}})
	=
	\varepsilon_{\mathrm{pre}}^{\mathrm{th}}, \quad d>0,
	\label{eq:thermal_inf}
\end{equation}
while $\varepsilon^{\mathrm{th}}(0)=0$ for all $\gamma$.
This corresponds to a strongly localized thermal strain near the kirigami bilayer boundary.

\medskip
Figure~\ref{fig:4}E illustrates the variation of the normalized thermal strain $\varepsilon^{\mathrm{\rm th}}/\varepsilon_{\rm pre}^{\mathrm{\rm th}}$ with normalized distance ${d}/{d_{\max}}$ for $\gamma=0.1$, 1, and $5$,
corresponding to characteristic length $L_c=d_{\max}/\gamma$ in Equation~\ref{eq:Lc}. 
As predicted by the asymptotic analysis, the response approaches a linear profile when $\gamma \to 0$, the case $\gamma=0.1$ is shown by black circular markers.
As $\gamma$ increases, the response becomes increasingly localized near the bilayer boundary (corresponding to $d/d_{\max}=0$), as shown by the blue and red markers at $\gamma=1$ and 5, respectively.
No markers are presented at $d/d_{\max}=0$ because the discrete mesh edge length is always greater than zero.
Figures~\ref{fig:4}F(i)-(iii) show the deformed configurations of pattern A with $\varepsilon_{\rm pre}^{\mathrm{\rm th}}=0.7$ for $\gamma=0.1$, 1, and 5, from top to bottom, respectively. Figure~\ref{fig:4}G shows the corresponding top views of the configurations shown in Figure~\ref{fig:4}F.
As $\gamma$ increases, the characteristic length $L_c$ decreases, leading to a more localized thermal strain field and consequently, more localized deformation, as illustrated in Figure~\ref{fig:4}F(iii) and Figure~\ref{fig:4}G(iii).

\medskip
Hereafter, we adopt the linear form for thermal strain field
\begin{equation}
    \varepsilon_i^{\mathrm{th}}
      = \varepsilon_{\mathrm{pre}}^{\mathrm{th}}\,
        \frac{d_i}{d_{\max}},
        \quad
    \label{eq:thermal_strain}
\end{equation}
which corresponds to the $\gamma \to 0$ limit of the exponential mode.
This choice serves as a first-order approximation that captures the experimentally observed spatial trend while avoiding the introduction of an additional length scale parameter. 
The varialbles $d_i$ and $d_{\max}$ are defined in Equation~\ref{eq:define_d},
and $\varepsilon_{\rm pre}^{\mathrm{th}}$ is the thermal strain that develops in a pure Shrinky Dink sheet.
}

\medskip
The planar strain mismatch at each hinge is computed as
\begin{equation}
    \Delta\varepsilon_j(\bm X, T) = \sum_{p=1}^4 s_p [\varepsilon_{jp}(\bm X, T)-\varepsilon^{\rm th}_{jp}(\bm X, T)],
    \label{eq:delta_eps}
\end{equation}
where $\varepsilon_{jp}$ and $\varepsilon^{th}_{jp}$ are the axial strains and axial thermal strains, respectively, of the four edges adjacent to hinge $j$ indicated by $1,2,3,$ and $4$ in Figure~\ref{fig:4}D, and $s_p \in \{+1, -1\}$ are orientation factors determined by the mesh geometry (see Figure~\ref{fig:4}D for edge numbering and orientation conventions).
Physically, $\Delta \varepsilon_j$ quantifies the difference in in-plane contraction between the two triangles sharing hinge $j$, which is the local geometric origin of natural curvature. The corresponding thermal strains, $\varepsilon_{jp}^{\mathrm{th}}(T)$, are computed from the experimentally measured contraction of the Shrinky Dink sheet at each temperature, as reported in Figure~\ref{fig:3}E.

\medskip
This reduced energy formulation incorporates both the spatial heterogeneity of the composite and the direct influence of external stimulus. By minimizing $E(\bm X, T)$ for a given $T$, we predict the equilibrium 3D shape for any kirigami design and actuation protocol. The model is validated in the next section by comparison with 3D scans of the fabricated structures.

{\color{red}
\subsubsection{Model Assumptions for the Adhesive Layer}
\medskip
We used a two-part Gorilla epoxy adhesive to bond the Shrinky Dink and PLA kirigami layers. 
After curing, the thickness of the adhesive layer is in the range of $(0.0,0.1]$ mm.
Recall that the thickness of the Shrinky Dink substrate is $h_1=0.3$mm, while the thickness of the PLA kirigami layer is in the range $h_2 \in [0.6, 1.0]$ mm for the different patterns considered. 
As a result, the total thickness of the bilayer composite lies in the range $[1.0,1.4]$ mm, and the adhesive layer accounts for less than 1/10 of the total thickness.
Given this small thickness ratio, the adhesive contribution to the overall bending and stretching stiffness of the composite is expected to be secondary. We therefore neglect the adhesive layer in the numerical model and represent the structure as an effective bilayer composite.

\medskip
If the adhesive were explicitly modeled as a third layer, reported material properties for cured epoxy adhesives indicate a Young’s modulus in the range 
1.5-4.0 GPa \cite{lapique_curing_2002, salom_mechanical_2018} at room temperature, and a Poisson's ratio between 0.33 and 0.43 \cite{lim_mechanical_2019}, according to studies of cured epoxy resin adhesives.
The corresponding shear modulus, $G=E/[2(1+\mu)]$, therefore lies in the range of 0.5-1.5 GPa at room temperature.
Under the thermal actuation temperature used in this study ($>300^\circ$F), which exceeds the glass transition temperature of epoxy adhesives, the cured adhesive undergoes pronounced thermal softening and exhibits a substantial reduction in stiffness, by two to three orders of magnitude~\cite{mccrum_principles_1997}.
Consequently, the Young's and shear moduli of the adhesive during thermal actuation are on the order of MPa, rather than GPa.

\medskip
Incorporating such a thin adhesive layer would effectively lead to a trilayer composite with slightly increased effective stretching and bending stiffnesses $k_s$ and $k_b$ in Equations~\ref{eq:ks_eff} and \ref{eq:D_eff}.
However, because the adhesive layer is thin compared to the composite, its primary effect would be a quantitative increase in stiffness rather than a qualitative change in deformation shape.

\medskip
Importantly, the focus of this work is on the coupling between in-plane stretching and out-of-plane bending induced by thermal strain mismatch in the composite. This coupling mechanism, as well as the resulting morphing behavior, is not sensitive to small variations in effective stiffness caused by the thin adhesive layer. We therefore neglect the adhesive layer in the numerical modeling without loss of generality, while noting that explicitly including it would not alter the main conclusions of the study.
}

\subsection{Numerical Results Validated with Experiments}
\label{sec:validation}

{\color{red}
\subsubsection{Model Parameters}
}

The bilayer composite is modeled as a single-layer elastic shell that incorporates thermal strains and dihedral angles as reference strains in the energy, Equation~\ref{eq:bilayer_energy}. 
Each region is characterized by its effective Young's modulus $Y$. 
For thin plates and shells,
it is reasonable to assume plane stress condition, under which the effective Young's modulus is given by $Y=Y_0/(1-\nu^2)$, where $Y_0$ is the original modulus value.
At room temperature, for the Shrinky Dink sheet, denoted as layer 1, $Y_{0,1}$ typically ranges from 0.8GPa to 3GPa~\cite{mishra_effect_2005}.
For the 3D-printed PLA layer, 
denoted as layer 2,
$Y_{0,2}$ ranges from 2GPa to 4GPa
\cite{crupano_investigating_2024}.
Without loss of generality, we take the Poisson's ratio as $\nu=0.33$ for both layers.
We set the original modulus $Y_{0,1}=0.8911$ GPa for the Shrinky Dink,
and $Y_{0,2}=2.6733$ GPa for the PLA layer,
giving effective moduli $Y_1=1$ GPa
and $Y_2=3$ GPa, respectively.
At a temperature over $300^\circ F$ which is required for thermal actuation,
the moduli decrease by two to three orders of 
magnitude compared to their values at room temperature \cite{lendvai_valorization_2023,tabi_application_2019,ricarte_tutorial_2024}.
So we set $Y_1=1$ MPa for the Shrinky Dink,
and $Y_2=3$ MPa for the PLA layer.
Given the scaling of stiffness and energy are both linear in Young's modulus, the unit from GPa to MPa does not affect the equilibrium solution.
The effective stretching stiffness for a bilayer composite is \cite{ugural_advanced_2011}
\begin{equation}
k_s=\sum_{i=1}^2 k_s^i,
\label{eq:ks_eff}
\end{equation}
where $k_s^i$ for each layer is given by Equation~\ref{eq:ks_kb}.

\medskip
The relation between flexural rigidity and bending stiffness of a plate is $k_b=(2/\sqrt{3})D$. For a homogeneous plate, $D=Y h^3/12$.
For a bilayer composite plate, we have the effective flexural rigidity
\begin{equation}
	D/b=D_{\rm eff}/b
	=\sum_{i=1}^2
	Y_i
	\biggl[
	\frac{h_i^3}{12}+h_i(y_i-\bar y)^2
	\biggr],
    \label{eq:D_eff}
\end{equation}
where the location of the bending neutral axis is
$\bar{y}= (\sum_{i=1}^2 Y_i A_i y_i)/(\sum_{i=1}^2 Y_i A_i)$ \cite{ugural_advanced_2011}.
We set the bottom of Shrinky Dink (layer 1) at \(y = 0\), so we have
$y_1 = h_1/2$, $y_2 = h_1 + h_2/2$,	$A_1 = b h_1$, and $A_2 = b h_2$.

\medskip
Recall that layer 1 in Equation~\ref{eq:D_eff} denotes the Shrinky Dink sheet, and layer 2 denotes the PLA layer.
In our experiments, we used thicknesses of
$h_1=0.3$mm for the Shrinky Dink sheet, and $h_2=0.7$ mm, $0.6$ mm, and $1.0$mm
for the PLA layer in patterns A, B, and C in Figure~\ref{fig:2}, respectively.

{\color{red}
\medskip
From our experimental observation, bending is activated more easily than in-plane stretching during the thermal actuation process. 
To improve numerical stability, we scale the effective stretching stiffness $k_s$ up by one order of magnitude.
This scaling is equivalent to reducing the effective bending stiffness $k_b$ by the same factor, since the elastic energy density in Equation~\ref{eq:bilayer_energy} depends linearly on both stiffnesses.
We have verified numerically that both approaches lead to identical equilibrium configurations.

\medskip
For each layer, Equation~\ref{eq:ks_kb} gives the ratio between stretching stiffness $k_s$ and bending stiffness $k_b$ as
\begin{equation}
\frac{k_s}{k_b} = 9 \frac{l^2_0}{h^2}.
\label{eq:ks_kb_ratio}
\end{equation}
For a thin shell with $l_0/h<1$, the bending stiffness is orders of magnitude smaller than the stretching stiffness.
As a result, in classical thin shell problems with free bending, stretching energy usually dominates bending energy due to the relatively large ratio $k_s/k_b$. 
However, in the present thermally actuated system, the deformation is governed by an imposed in-plane strain mismatch. 
As the structure bends to accommodate the prescribed thermal strain, the
stretching term contributing to the stretching energy, $\varepsilon-\varepsilon^{\rm th}$, remains small, and the dominant contribution arises from bending. 
This behavior is characteristic of curvature-dominated responses.

\medskip
We choose to scale $k_s$ rather than $k_b$ because stretching contributes less to the total strain energy than bending for the geometries and boundary conditions considered here.
We have verified that scaling $k_s$ upward or scaling $k_b$ downward by the same factor leads to identical solutions.
However, when $k_s$ is scaled by smaller factors, such as 2 or 5, some regions may favor stretching over bending, leading to different equilibrium configurations. Scaling $k_s$ by one order of magnitude consistently generates stable solutions.
}

{\color{red}
\medskip
We solve for equilibrium at each time step using a Newton-Raphson iteration, which corresponds to locally minimizing the total potential energy under the prescribed boundary conditions. As is common for solvers of nonlinear mechanical systems, the Newton method converges to a nearby stable equilibrium configuration rather than guaranteeing convergence to a global minimum.}
Thermal actuation is applied in incremental load steps with an adaptive step size chosen to satisfy the convergence criterion $|\Delta \bold{X} | / |\bold{X}|<10^{-5}$. 
Thermal strains are prescribed according to Equation~\ref{eq:thermal_strain}, see Section~\ref{sec:energy} for justification.
Because the Shrinky Dink layer contracts on heating, the induced thermal strains are negative in sign, $\varepsilon^{\rm th}_{\rm pre}<0$. 
{\color{red}
To account for practical oven heating and to improve numerical stability, we introduce a small, spatially uniform thermal flux load during the heating process.
This load gradually decays to zero by the end of actuation, consistent with the sample being removed from the oven. The small load promotes smooth convergence from the initial flat configuration to a physically meaningful deformation branch, especially in cases where the structure may initially bifurcate to bend either upward or downward.

\medskip
In experiments, bilayer composites are observed to occasionally bend upward or downward depending on small and often imperceptible manufacturing imperfections. Importantly, however, the resulting 3D shapes are very similar. Our numerical simulations reproduce stable 3D configurations for a given composite design and thermal actuation process using the current energy minimization approach. The close quantitative agreement between simulations and experiments, with SSIM values above 0.8, indicates that the computed final shapes approximate the experimental configurations.

\medskip
While a rigorous proof that the numerical solution corresponds to the global minimum is generally not available for nonlinear shell systems with multiple stable equilibrium states, the good consistency with experimental observations suggests that the calculated configurations represent the experimental responses.
}

\medskip
We discretize the composite domain into triangular elements, three meshes in the reference configuration are shown in Fig.~\ref{fig:4}. We have verified the finite-difference approximations of the gradient and Hessian of the total energy. The numerical implementation was further verified with analytical solutions for uniaxial tension and compression of a simple geometry domain, and the deflection of a cantilever beam under an endpoint load. Complex behaviors, such as bulging and necking also emerge from simple deformation modes \cite{zhang_iterated_2024}. 

{\color{red}
\medskip
To check the dependence of the results on mesh resolution, we performed a mesh sensitivity analysis using three different meshes for each pattern.
For example, for pattern A, the composite was discretized using meshes with 863, 970, and 1080 nodes.
The mean structural similarity index (SSIM, defined in Section~\ref{sec:ssim}) between the deformed configurations obtained using the 863- and 970-node meshes is 0.97, and increases to 0.99 between meshes with 970 and 1080 nodes, indicating convergence of the solution with respect to the mesh discretization. Similar convergence was observed in nodal displacement fields. Based on these results, the meshes shown in Figure~\ref{fig:4} A-C(i) are used in simulations.
}
Given these verifications, all subsequent simulations are obtained with the same numerical implementations.

{\color{red}
\subsubsection{Validation of DES Results with Experimental Scans}
}

\begin{figure}[!htb]
	\begin{center}
		{\includegraphics[width=6.5in]{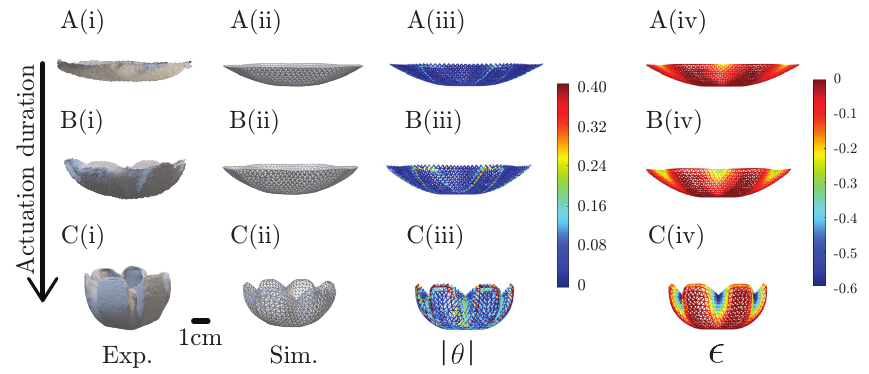}}
		\caption{Representative snapshots,  starting with the pattern of Figure~\ref{fig:2}A during the thermally actuated manufacturing process.
		From the top to the bottom rows, the duration of the thermal stimulus  increases. In each column, (i) shows the experimentally scanned shape, (ii) shows the numerical simulation, (iii) shows the distribution of the absolute value of the dihedral angle $|\theta|$, and (iv) shows the distribution of the axial strain $\varepsilon$.}
		\label{fig:5}
	\end{center}
\end{figure}

\textbf{Figure~\ref{fig:5}} presents the experimental scans, numerical simulations,  contours of dihedral angle $|\theta|$, and axial strains $\varepsilon$ for a circular composite with a six-arm PLA pattern, Figure~\ref{fig:2}A.
The term ``stimulus'' refers to the monotonically increasing stimulus, which is heat here, during the process. From top to bottom, the simulations are in equilibrium with prescribed thermal strains $\varepsilon^{\rm th}_{\rm pre} = -0.20, -0.30$ and $-0.77$, respectively.
As noted above, these thermal strains are nonuniform and vary with
each material point's distance to the PLA pattern.
Larger thermally induced shrinkage results in more significant deformations and
axial strains. A greater mismatch in axial strains leads to larger dihedral angles.
To accommodate these deformations, the structure morphs from a planar configuration into a 3D shape.
Deformation localizes heavily between the arms of the 3D-printed PLA pattern, where the Shrinky Dink layer undergoes the largest shrinkage,
indicated by A-C(iii) and A-C(iv) in Figure~\ref{fig:5}.
Overall, excellent agreement is observed between the scanned experimental shapes (first column) and the numerical simulations at each stage of the thermal actuation.

\begin{figure}[!htb]
	\begin{center}
		{\includegraphics[width=6.5in]{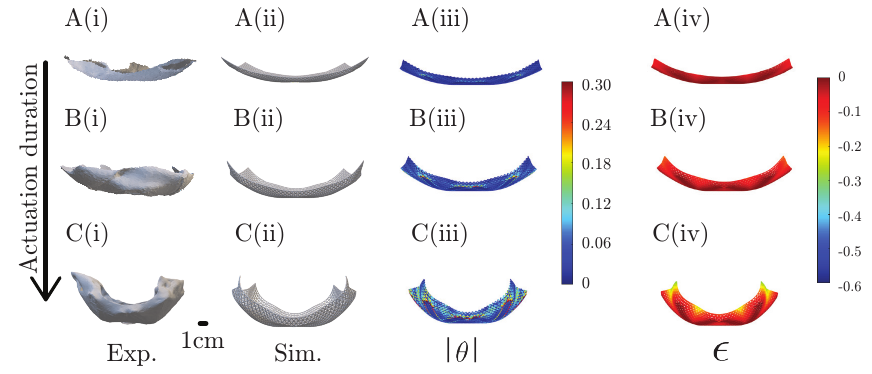}}
		\caption{Representative snapshots starting with the pattern of Figure~\ref{fig:2}B during the thermal actuated manufacturing process.
			Description of the subfigures is available in the caption of Figure~\ref{fig:5}.}
		\label{fig:6}
	\end{center}
\end{figure} 

\medskip
\textbf{Figure~\ref{fig:6}} shows
the results for the rectangular composite with a two-star PLA pattern in Figure~\ref{fig:2}B, subjected to thermal strains $\varepsilon^{\rm th}_{\rm pre} = -0.08, -0.15$ and $-0.30$, from the top to the bottom, respectively.
As the prescribed thermal axial strain increases over time, the structure develops larger axial strains $\varepsilon$, strains mismatch, dihedral angles $|\theta|$, and overall curvature.
Deformation localizes near the four corners of the Shrinky Dink layer.
The slight asymmetry in the deformed shapes arises from the mesh used in the numerical simulation and minor anisotropies in the bilayer composite due to physical manufacturing defects.
Nevertheless, the numerical simulations closely approximate the scanned experimental results.

\begin{figure}[!htb]
	\begin{center}{\includegraphics[width=6.5in]{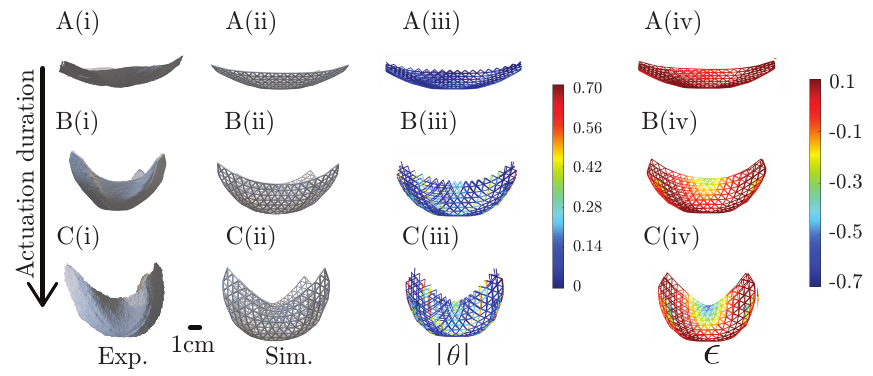}}
		\caption{Representative snapshots starting with the pattern of Figure~\ref{fig:2}C during the thermal actuated manufacturing process.
			Description of the subfigures is available in the caption of Figure~\ref{fig:5}.}
		\label{fig:7}
	\end{center}
\end{figure} 

\medskip
Similarly, \textbf{Figure~\ref{fig:7}} illustrates the deformed configurations at thermally actuated strains $\varepsilon^{\rm th}_{\rm pre} = -0.10$, $-0.30$ and $-0.50$ for the cross pattern in Figure~\ref{fig:2}C. As the prescribed thermal strain increases, deformation concentrates increasingly between pattern's arms. Because the Shrinky Dink layer is square rather than circular, these concentration zones form four sharp ``petals'' instead of the rounded bowl seen in Figure~\ref{fig:5}.

\medskip
These three examples demonstrate excellent agreement between scanned experimental 3D shapes and the numerical simulations morphed from 2D planar composites across various domain geometries, patterns, and levels of thermal actuation.
More complex structures can be achieved by assembling multiple shapes and using multiscale modeling.
The substantial computational demand of  multiscale modeling across multiple time and length scales can be mitigated through iterated learning of operators for out-of-distribution data \cite{zhang_iterated_2024}.
Potential applications of the three examples include deployable flower petals, bowls, frame-based pillows, canoes, and tent-type structures, each achieved by selecting an appropriate reference length scale.

{\color{red}
\captionsetup{labelfont={color=red},textfont={color=red}}
\medskip
The current model predicts deformation morphology under the assumption of ideal bonding and purely elastic behavior. However, during experiments we observed localized failure phenomena, including interface debonding near the boundary between the kirigami PLA layer and the Shrinky Dink layer, as well as strain localization at sharp corners. 
Because the thermal actuation temperature exceeds $300^\circ$F (shown by Figure~\ref{fig:3}E), which is higher than the glass transition temperatures of both the Shrinky Dink substrate and the PLA layer, the dominant deformation mechanisms associated with local failures are likely creep deformation, stress relaxation, and viscoelastic effects, rather than brittle fracture or rate-independent yielding. In the absence of significant external loading that would cause tearing, ductile fracture was not observed.

\medskip
The localized failure phenomena could potentially explain the slight asymmetry in deformation observed experimentally in Figures~\ref{fig:5}, \ref{fig:6}, \ref{fig:7} A(i), B(i) and C(i). At present, the model does not capture such failure mechanisms. Nevertheless, the framework can be extended to incorporate inelastic effects, for example by introducing Perzyna viscoplasticity based on internal variable theory~\cite{zhang_iterated_2024}, as well as power-law creep models to account for time-dependent creep deformation and stress relaxation~\cite{zhang_identification_2021}.

\medskip
\subsubsection{Young's Modulus and Nominal Stiffness Before and After Thermal Actuation}

\medskip
We further estimate the Young's modulus of the Shrinky Dink layer before and after thermal actuation. 
Given the small thickness (approximately $0.3$ mm) of a Shrinky Dink sheet before actuation, we consider a thin cantilever beam with one end fixed with all DOFs constrained, and the other end free. With gravity acting as a distributed load on the beam, the classical relation between
the distributed load $q$ and the free-end deflection $\delta_0$ is given by
\begin{equation}
    \delta_0 = \frac{q L^{4}}{8 E I_0}, \quad
    I=b_0h_0^3/12, \quad
    q = \rho g b_0 h_0,
  \label{eq:exp_beam1}
\end{equation}
where $b_0$ is the width, $h_0$ is the thickness, and the density is $\rho=1.05\rm{g/cm}^3$ before thermal actuation.
Using this relation, we estimate the Young's modulus as $E=2.60$ GPa with a standard deviation of $0.32$ GPa before thermal actuation.

\medskip
After thermal actuation, the Shrinky Dink sheet becomes thicker, with a thickness of approximately $1.8$ mm, and gravity no longer generates a noticeable deformation. In this case, we apply a point load at the free end of the cantilever beam, and measure the applied force using a Rokubi six-axis force sensor (BOTA BFT-ROKA-SER-M8) \cite{BotaSystems_FT_Manual}.
Using the classical relation
\begin{equation}
    \delta_a = \frac{P L^{3}}{3 E I_a}, \quad
    I_a=b_ah_a^3/12
  \label{eq:exp_beam2}
\end{equation}
where the subscript $(\cdot)_a$ denotes values after thermal actuation,
we estimate the Young's modulus as $E=2.65$ GPa with a standard deviation of $1.58$ GPa after thermal actuation. The relatively large standard deviation is due to the high nominal stiffness of the thickened Shrinky Dink layer after thermal actuation, which makes accurate deformation measurements more challenging. The reported results are averaged over five samples in each case.

\medskip
The nominal stiffness of a complex composite is an important structural performance metric, particularly in applications such as aerospace engineering. 
The nominal stiffness $K_{\rm nom}$ of a structure is defined as the ratio between the applied generalized load $P_{\rm nom}$ and the measured generalized displacement $\delta_{\rm nom}$.
For a cantilever beam with a point load $P$ applied at the free end, leading to a displacement $\delta_a$, we set $P_{\rm nom,a}=P$, $\delta_{\rm nom,a}=\delta_a$, and the nominal stiffness is given by 
\begin{align}
    K_{\rm nom,a}=P_{\rm nom,a}/\delta_{\rm nom,a}=3EI_a/L^3,
    \label{eq:stiffness_nominal1}
\end{align}
from Equation~(\ref{eq:exp_beam2}).
For a cantilever beam subjected to a distributed load $q$ with free-end displacement $\delta_0$, we set $P_{\rm nom,0}=qL$ and $\delta_{\rm nom}=\delta_0$, which gives
\begin{align}
    K_{\rm nom,0}=P_{\rm nom,0}/\delta_{\rm nom,0}=8EI_0/L^3,
  \label{eq:stiffness_nominal}
\end{align}
from Equation~(\ref{eq:exp_beam1}).
Here, the moment of inertia $I$, which changes significantly due to geometric thickening after thermal actuation, dominates the nominal stiffness. 
Using
$I_0~10^{-14}m^4$ and $I_a~10^{-12}m^4$, along with
$L_0=0.16$ m and $L_a~0.073$ m, we obtain
$K_{\rm nom,a}=142 \pm 52$ N/m, and
$K_{\rm nom,0}=0.2357 \pm 0.0016$ N/m.
It is worth noting that the estimated Young's modulus values for the Shrinky Dink material before and after thermal actuation remain close at approximately 2.6 GPa, whereas the corresponding nominal stiffness values differ by more than a factor of 600.

\medskip
These results indicate that, although the Young's modulus of the Shrinky Dink material remains essentially unchanged before and after thermal actuation, the rigidity of the cooled, permanently deformed structure increases dramatically due to geometric stiffening, resulting in a nominal stiffness more than two orders of magnitude larger than that of the undeformed configuration.
}

\subsection{Structural Similarity Index Measure}
\label{sec:ssim}

To compare the scanned experimental shapes with numerically simulated ones quantitatively, 
we use the structural similarity index measure (SSIM), defined at each voxel in a 3D domain.
For completeness, we briefly describe our application.
We discretize the 3D domain into an $N_s \times N_s \times N_s$ grid and
calculate SSIM at each voxel as \cite{nilsson_understanding_2020}
\begin{equation}
	\text{SSIM} =
	\frac{\bigl(2\,\mu_1\,\mu_2 + c_1\bigr)\,\bigl(2\,\sigma_{12} + c_2\bigr)}
	{\bigl(\mu_1^2 + \mu_2^2 + c_1\bigr)\,\bigl(\sigma_1^2 + \sigma_2^2 + c_2\bigr)},
\end{equation}
where $N_s=10$.
The voxel intensities are normalized to $[0, L_s]$, 
and we use the default parameter values
$k_1 = 0.01$, $k_2 = 0.03$, $L_s=1$, and $c_i = (k_i*L_s)^2$ for $i=1,2$ \cite{nilsson_understanding_2020}.
Here, $\mu_1$, $\mu_2$ are local means,
$\sigma_1^2$, $\sigma_2^2$ the variances, and $\sigma_{12}$ the covariance.
To make the scanned and simulated shapes comparable,
we uniformly scale, translate, and rotate the simulated shapes to
align with the scanned experimental data.
For our results, the SSIM ranges from 0 to 1, with 0 indicating no structural correlation and 1 indicating perfect structural similarity.

\begin{figure}[h!]
	\centering
	\includegraphics[width=6.5in]{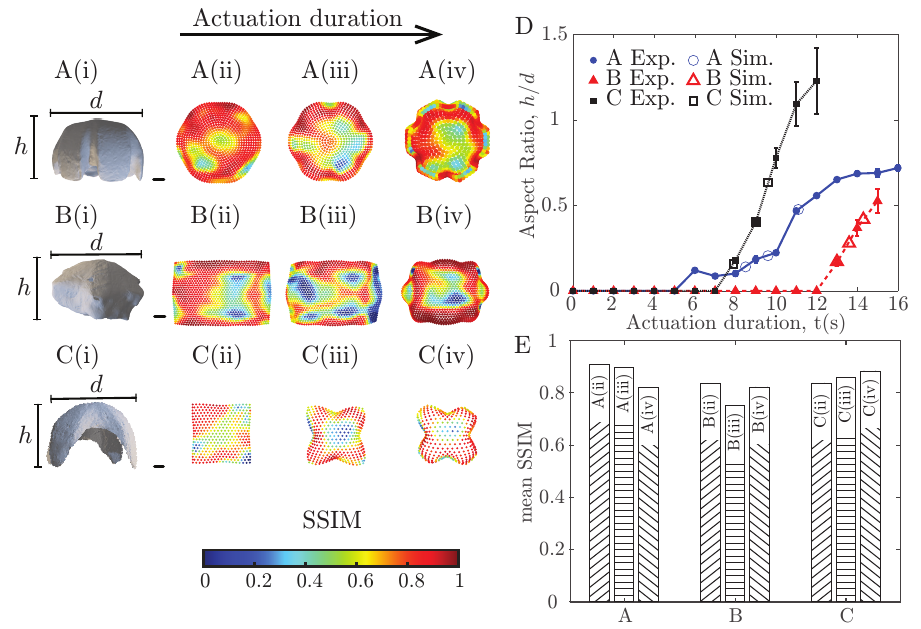}
	\caption{Validation of numerical simulations against experimental results using SSIM.
    A-C(i) show side views with the definition of the aspect ratio $h/d$.
    A-C(ii-iv) present top-view SSIM contours at three morphing stages for each pattern.
    D shows the evolution of $h/d$ during thermal actuation, where solid markers denote experimental means and hollow markers denote numerical simulations.
    E summarizes the mean SSIM values for the contours in A-C(ii-iv).}
	\label{fig:8}
\end{figure}

\medskip
\textbf{Figure~\ref{fig:8}} summarizes the SSIM between experimental scans and numerical simulations for all three composite shapes and their morphing processes.
Panels (i) in A-C show a side view of each shapes alongside definition of the aspect ratio $h/d$.
Panels (ii)-(iv) present top-view SSIM contours
at three morphing stages for each geometry through A-C.
The nine morphing stages are corresponding to the nine stages in Figures~\ref{fig:5}-\ref{fig:7}, respectively.
Despite anisotropies in the Shrinky Dink sheet, glue effects, sharp cut boundaries, and inevitable scanning noise, the SSIM contours show close agreement between simulation and experiment, with most regions appearing in yellow to red.

\medskip
In Figure~\ref{fig:8}D, 
solid markers with error bars indicating standard deviations show the evolution of the aspect ratio $h/d$ over time during thermal actuation.
The entire morphing process completes in approximately 15 seconds, dramatically faster than conventional 3D-shape fabrication methods, which often require minutes to hours.
Each pattern exhibits a slightly different morphing timeline.
The six-arm star on the circular sheet (pattern A) begins deforming earliest at around 5s, and its $h/d$ plateaus after 12s.
The two-star pattern on the rectangular sheet (pattern B) begins to morph latest at around 12s.
The cross pattern on the square sheet (pattern C) starts at 7s, and ultimately reaches an aspect ratio of $h/d>1.2$.
Hollow markers denote the numerical simulations obtained by prescribing thermal strains as described in Sections~\ref{sec:energy} and \ref{sec:validation}.
Figure~\ref{fig:8}E plots the mean SSIM values for the nine SSIM contour plots shown in panels (ii)-(iv).
The mean SSIM is given by
$\overline{\mathrm{SSIM}}
= \tfrac{1}{N_s^3}
\sum_{i=1}^{N_s}
\sum_{j=1}^{N_s}
\sum_{k=1}^{N_s}
\mathrm{SSIM}(i,j,k)$.
All mean SSIM values are around 0.8, indicating close agreement between the numerical simulations and the scanned experimental shapes.

{\color{red}
\medskip
The restriction on the choice of actuation duration (16 s, 15 s, and 12 s for patterns A to C, respectively) is that it must fall within an appropriate range. In experiments, significantly longer heating times would lead to excessive material softening or partial melting of the composite, resulting in uncontrolled deformation. For the materials used in this study, thermal actuation durations within approximately 20 s remain mechanically stable. Accordingly, we select actuation durations that yield visually representative morphing configurations and clearly demonstrate the design capability, while avoiding excessive softening.
With longer heating times, both the Shrinky Dink substrate and the PLA components soften significantly. Due to gravity and the reduced stretching and bending stiffnesses, the structures may fold inward or collapse, transitioning beyond the elastic deformation regime considered in our model. Since the present model focuses on elastic thermo-mechanical coupling, samples are removed from the oven before significant softening or melting occurs.

\medskip
Importantly, the objective of the thermal actuation is not to reach a steady state. The horizontal axis in Figure~\ref{fig:8}D represents the prescribed actuation duration rather than physical time in a dynamic sense. Thermal actuation can be readily controlled by removing the sample from the oven and allowing it to cool to room temperature, at which point deformation effectively ceases. As a result, reaching a steady state is not required for either the experiments or the simulations.

\medskip
The different evolution rates of the aspect ratio $h/d$ among patterns arise partially from geometric effects. Composites with patterns B and C contain sharper corners than pattern A, which contribute more noticeably to out-of-plane deformation during actuation. Consequently, the aspect ratio in patterns B and C are more sensitive to out-of-plane deformation than in pattern A under the same thermal actuation process.
}

{\color{red}
\captionsetup{labelfont={color=red},textfont={color=red}}

\subsection{Numerical Results Verified with Finite Element Simulations}
\label{sec:FEA}

We further verify the results from discrete elastic shell (DES) model by comparison with finite elements (FE) simulations performed using Abaqus \cite{dassault_systemes_simulia_corp_abaquscae_2020}. 
The bilayer composite is modeled as two shell layers, each discretized with S3R shell elements.
Each layer contains $N$ nodes, resulting in a total of $2N$ nodes.
The two layers are coupled through a tie constraint between the contacting surfaces. 
Both layers are modeled as isotropic Mooney-Rivlin hyperelastic materials with strain energy density
\cite{dassault_systemes_simulia_corp_abaquscae_2020}
\begin{equation}
	W(\mathbf F)
	=
	C_1(\bar I_1-3)
	+
	C_2(\bar I_2-3)
	+
	\frac{1}{D_1}(J-1)^2,
	\label{eq:MR}
\end{equation}
where $\bar I_1$ and $\bar I_2$ are the first and second invariants of the isochoric right Cauchy-Green tensor, respectively, $J$ is the determinant of the deformation gradient, and $C_1$, $C_2$ and $D_1$ are fitting parameters for materials. 

\medskip
In the infinitesimal strain limit, the Mooney-Rivlin model can be linearized to an equivalent isotropic elastic solid characterized by shear modulus $G$ and bulk modulus $K$ with
\begin{equation}
	C_1 = C_2 = \frac{G}{4},
	\qquad
	D_1 = \frac{2}{K},
\end{equation}
where the derivation is provided in Appendix~\ref{app:MR} under the assumption $C_1 = C_2$.

\medskip
Recall that during thermal actuation,
the Young's moduli are taken as $Y_{0,1}=0.8911$ MPa for the Shrinky Dink substrate layer and $Y_{0,2}=2.6733$ MPa for the PLA kirigami layer, reduced from their room temperature values as discussed in Section~\ref{sec:numerics}.
The Poisson's ratio is $0.33$ for both materials.
This results in $C1=C2=8.38 \times 10^4$Pa and $D_1=2.29 \times 10^{-6}\rm{Pa}^{-1}$ for Shrinky Dink polystyrene, 
and $C1=C2=2.51 \times 10^5$Pa and $D_1=7.63 \times 10^{-7}\rm{Pa}^{-1}$ for the PLA layer.

\medskip
We modeled the deformation using an explicit time integration in Abaqus~\cite{dassault_systemes_simulia_corp_abaquscae_2020}. The mass density is set to $1.05\rm{g/cm}^3$ for the Shrinky Dink Polystyrene substrate layer \cite{ashby_materials_2012}, and $1.25\rm{g/cm}^3$ for the kirigami PLA layer \cite{khouri_polylactic_2024}.
The thickness of the Shrinky Dink layer is 0.3mm. The thickness of PLA layer is 0.7mm for pattern A, 0.6 mm for pattern B, and 1.0mm for pattern C.
The kirigami layer has a zero thermal expansion coefficient, whereas the substrate layer has a nonzero thermal expansion coefficient. Each node in the FE model has 6 DOFs. 

\medskip
\textbf{Figure~\ref{fig:9}} shows the verification of the DES model against FE simulations for two representative cases.
The first row (A) corresponds to pattern A in Figure~\ref{fig:4} with thermal strain parameter $\varepsilon^{\rm th}_{\rm pre}=-0.77$,
and the second row (B) corresponds to pattern C with $\varepsilon^{\rm th}_{\rm pre}=-0.50$.
The first column shows the deformed configuration, 
and the second column shows the contours of the minimum principal logarithmic strain $\varepsilon^{\rm LE}_{\min}$, 
the FE simulations closely reproduce the experimentally scanned configurations and show good agreement with DES results in 
Figures~\ref{fig:5} and \ref{fig:7}.
The last column shows the top-view SSIM contours computed between the FE and DES results. The mean SSIM values are 0.83 and 0.81 for cases A(iii) and B(iii), respectively,
demonstrating quantitative agreement between the DES and FE simulations.

\begin{figure}[h!]
	\centering
	\includegraphics[width=6.0in]{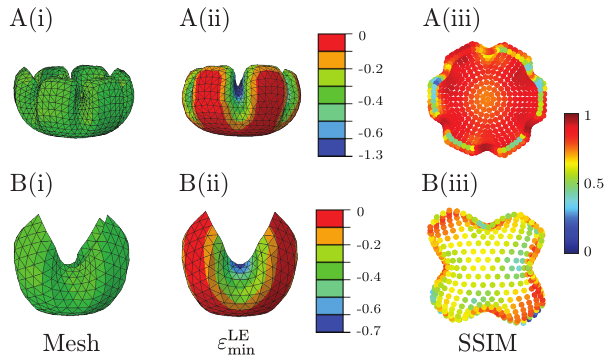}
	\caption{Verification of discrete elastic shell (DES) simulations against finite element (FE) results using Abaqus. 
    The first row (A) shows results for pattern A with thermal strain parameter $\varepsilon^{\rm th}_{\rm pre}=-0.77$, 
    and the second row (B) shows the results for pattern C with $\varepsilon^{\rm th}_{\rm pre}=-0.50$.
    Column (i) shows the deformed configurations, column (ii) shows contours of the minimum principal logarithmic strain $\varepsilon^{\rm LE}_{\min}$,
    and column (iii) shows the top-view SSIM contours computed between the FE and DES results. The mean SSIM values are 0.83 and 0.81 for A(iii) and B(iii), respectively.}
	\label{fig:9}
\end{figure}

\medskip
For the bilayer composite considered here, the FE model represents the structure as two shell layers coupled by a tie constraint. Each layer contains 
N nodes connected by S3R shell elements, resulting in a total of 
2N nodes with six degrees of freedom per node. Both layers are modeled as Mooney-Rivlin hyperelastic materials.
In contrast, the DES method models the bilayer composite as a reduced single layer shell with N nodes and 3 DOFs per node, using the elastic energy in Equation~\ref{eq:bilayer_energy}. 
Using the same numerical mesh and temperature change (i.e., thermal strain) distribution, the CPU time required to complete the simulation is 738 s for the FE model and 420 s for the DES model for pattern A, and 218 s for the FE model and 183 s for the DES model for pattern C, on a personal computer\footnote{A Dell Inspiron 3891 with an Intel Core i7-10700 CPU (8 cores) at 2.90 GHz processor, 12 GB of memory, a 512 GB SSD, and a 2 TB HDD (7200 RPM).}.
The improved computational efficiency of the DES approach arises from the reduced number of DOFs and numerical acceleration techniques, including sparse matrix assembly and sparse LU factorization (via the splu function in SciPy) at each Newton iteration.
}

\section{Conclusion}

We introduced a new energy formulation 
for reduced-order modeling of composites as a single layer material, enabling efficient simulation of stimulus-driven morphing across a broad range of systems beyond bilayers.
The formulation couples out-of-plane bending with in-plane stretching, expressed as functions of the applied stimulus, allowing accurate modeling of permanent and stimulus-induced deformations.

\medskip
We also developed a robust fabrication protocol to transform planar composites into 3D structures via thermal actuation, achieving a repeatable success rate exceeding 90\%. 
Three representative geometries --
a six-arm star on a circle, a two-star on a rectangle, and a cross on a square -- showed excellent agreement between
numerical simulations and experiments across various morphing stages, as evidenced by both qualitative comparisons in Figures \ref{fig:5}-\ref{fig:7},
and quantitative SSIM values in Figure~\ref{fig:8}.

\medskip
The close agreement between simulations from our reduced-order model and  experiments highlights the predictive power and broad potential for fast, low-cost, and precise design and manufacturing of complex structures through emergent morphogenesis.

\section{Experimental Section}

{\it Materials and Methods}:
We used commercially available Shrinky Dink polystyrene sheets (Cridoz Shrinky Art Paper Sheets) and 3D-printed PLA kirigami inspired patterns. The kirigami components were designed using Fusion 360 and SolidWorks CAD software \cite{dassault_systemes_solidworks_corp_solidworks_2025}, utilizing Python scripts to rapidly assign and vary geometry. The designs were fabricated with a fused deposition modelling 3D printer (Bambu Lab 1PS) using PLA filament chosen for its thermal resistance during oven experimentation and its mechanical properties. Adhesion between layers was achieved using a two-part  \emph{Gorilla} epoxy adhesive, selected for its bond strength, thermal stability, cost, and ease of use.

\medskip
{\it Shrinky Dink Properties}:
Shrinky Dink is a thermoplastic polystyrene polymer that exhibits isotropic in-plane shrinkage when heated. We define the shrink ratio as
$L/L_0$,
where $L$ is the instantaneous length and $L_0$ is the original length. Figure~\ref{fig:3}E shows the variation of normalized length $\bigl(L/L_0\bigr)$ with normalized temperature $T/T_g$. The shrink ratio decreases with increasing temperature and plateaus once $T/T_g$ exceeds 1.15, indicating isotropic behavior as the shrink ratios along orthogonal in-plane directions are similar. Above $300\,^\circ\mathrm{F}$, the $x$- and $y$-axis gauge lengths coincide and the standard deviation bars contract, reflecting highly reproducible shrinkage. To ensure uniform and predictable actuation, all subsequent experiments were conducted within this temperature regime.

\medskip
{\it Manufacturing Protocol}:
The fabrication followed four key steps:
\begin{enumerate}
	\item Print the 3D kirigami design on a fused deposition modeling printer.
	\item Bond the 3D printed kirigami elements to pre-cut Shrinky Dink sheets with a thin, uniform layer of epoxy.
	\item Cure under distributed weight $(6.8\times10^2\ \mathrm{N/m^2})$ over 10 h at room temperature to ensure full adhesion.
	\item Thermally activate the composite in a laboratory oven preheated to $300\,^\circ\mathrm{F}$. Each sample was placed with the Shrinky Dink side facing down on a non-stick foil base and heated for 15s. The 3D morphing process was observed visually, and samples were removed by hand immediately upon full shape transformation.
\end{enumerate}

\medskip
{\it Pattern Design and Results}:
Morphing samples were generated via a Python script that produced random polygonal kirigami patterns on a 2D domain. Visual inspection of the resulting 3D shapes revealed that local curvature propagated along the printed line directions. In a parametric study, we varied three parameters: line width, PLA layer thickness, and line count per polygon. Empirically, increasing line count yielded smoother and more continuous circumferential curvature, whereas fewer lines produced segmented outlines. Thicker and wider PLA lines resisted actuation, requiring longer oven exposure due to higher bending and stretching stiffness. Consequently, designs with more than four lines were adopted for rapid, reliable thermal response and consistent circumferential curvature across samples.
Three planar patterns were therefore designed to benchmark the aforementioned observations and to investigate the numerical accuracy of simulations in comparison to the experimentally fabricated patterns. They are:
A. \textbf{Single-star}: Fabricated by intersecting three coplanar ``arms'', each \(90\,\mathrm{mm}\) long, at \(60^\circ\) increments through the centroid, generating six \(45\,\mathrm{mm}\) radial arms.
B. \textbf{Double-star}: Superimposed two such stars aligned coaxially, separated by a \(56.73\,\mathrm{mm}\) spine, giving an overall tip-to-tip length of \(146.7\,\mathrm{mm}\).
C. \textbf{Orthogonal X-cross}: Incorporating four arms spaced at \(90^\circ\) intervals, each extending \(56.7\,\mathrm{mm}\) from the center of the shape and spanning \(99.99\,\mathrm{mm}\) along the principal diagonal, with each arm chamfered at its end via the addition of a right-angled triangle, \textbf{Figure~\ref{fig:9}}.

\medskip
{\it Computational Reconstruction}:
The computational reconstruction utilized Polycam photogrammetry software alongside a high-resolution camera and a custom-printed stand to capture 80 images around each sample from multiple angles. These images were processed in Polycam to generate and edit 3D models in STL and OBJ formats. To enhance reconstruction consistency and spatial accuracy, we applied fiduciary markers to the samples prior to imaging, facilitating reliable feature recognition and reducing computation time \cite{surmen_photogrammetry_2023}. The resulting  meshes were scaled to real-world dimensions based on known physical measurements and subsequently 3D-printed to validate scale fidelity against experimental results.

\medskip
{\it Aspect Ratio}:
To extract the mean aspect ratio of each experimental sample, we conducted a frame-by-frame analysis at 1 s intervals using a high resolution video of the thermal actuation procedure. Frames were imported into ImageJ \cite{rasband_imagej_2025}, where a threshold mask isolated the sample from the background for accurate perimeter and geometry detection. The straight-line tool was used to manually define key dimensions, whose pixel lengths were recorded. Conversion from pixels to physical units was performed using a known reference length inside the oven. From these measurements, in-plane shrinkage and out-of-plane height expansion were computed and expressed as the numerical aspect ratio $h/d$, as demonstrated in Figure~\ref{fig:8}.

\medskip
\textbf{Supporting Information} \par 
Supporting Information is available from the Wiley Online Library or from the author. The data and codes are available on GitHub:
\url{https://github.com/StructuresComp/discrete-shells-shrinky-dink}

\medskip
\textbf{Acknowledgments}
This work was supported by Samsung Electronics Co., Ltd., and the National Science Foundation under Grants CMMI-2332555, CAREER-2047663, and CMMI-2209782.

\medskip
\textbf{Conflict of Interest}
The authors declare no conflict of interest.

\medskip
\textbf{Author Contributions}
Y.Z. and A.A. contributed equally to this work. 
Y.Z. developed the numerical methodology.
Y.Z., A.A., and M.K.J. designed the experiments.
A.A. performed the majority of the experiments and collected most of the data.
Y.Z. also contributed to experiments and  data collection.
Y.Z., A.A., and M.K.J. performed conceptualization, investigation, and formal analysis.
M.K.J. acquired funding, and supervised the project.
Y.Z., A.A., and M.K.J. wrote, reviewed, and edited the manuscript.
All authors read and approved the final manuscript.

\medskip
\textbf{Data Availability Statement}
The data that support the findings of this study are available from the
corresponding author upon reasonable request.

{\color{red}

\section{Appendix A. Asymptotic Analysis of Thermal Strain Distribution}
\label{app:limit}

We assume an exponentially varying thermal strain field
\begin{equation}
	\varepsilon^{\mathrm{th}}(d)
	=
	\varepsilon_{\mathrm{pre}}^{\mathrm{th}}
	\,
	\frac{1 - \exp\!\left(-{d}/{L_c}\right)}
	{1 - \exp\!\left(-{d_{\max}}/{L_c}\right)},
	\label{eq:thermal_exp_dimensional}
\end{equation}
where $d \in [0,d_{\max}]$ denotes the distance from an edge midpoint
to the closet kirigami pattern boundary,
$L_c>0$ is a characteristic decay length controlling the spatial localization,
and $\varepsilon_{\mathrm{pre}}^{\mathrm{th}}$ is the prescribed thermal strain parameter.
This ensures
$\varepsilon^{\mathrm{th}}(0)=0$ and
$\varepsilon^{\mathrm{th}}(d_{\max})=\varepsilon_{\mathrm{pre}}^{\mathrm{th}}$.

We define a dimensionless localization parameter $\gamma \in (0, \infty)$ to relate $L_c$ with $d_{\max}$
\begin{equation}
	L_c = {d_{\max}}/{\gamma},
\end{equation}
Equation~\eqref{eq:thermal_exp_dimensional} can be expressed in nondimensional form as
\begin{equation}
	\varepsilon^{\mathrm{th}}(d/d_{\max})
	=
	\varepsilon_{\mathrm{pre}}^{\mathrm{th}}
	\,
	\frac{1 - \exp(-\gamma d/d_{\max})}
	{1 - \exp(-\gamma)}.
	\label{eq:thermal_exp_dimensionless}
\end{equation}

In the limit $\gamma \to 0$, a Taylor expansion of the exponential term is
\begin{equation}
	\exp(-\gamma  d/d_{\max}) = 1 - \gamma  d/d_{\max} + O(\gamma^2).
\end{equation}
Then Equation~\eqref{eq:thermal_exp_dimensionless} reduces to
\begin{equation}
	\lim_{\gamma \to 0}
	\varepsilon^{\mathrm{th}}(\frac{d}{d_{\max}})
	=
	\varepsilon_{\mathrm{pre}}^{\mathrm{th}}
	\frac{d}{d_{\max}},
	\label{eq:thermal_linear}
\end{equation}
which corresponds to a linear thermal strain distribution.

On the other hand, in the limit $\gamma \to \infty$,
Equation~\eqref{eq:thermal_exp_dimensionless} reduces to
\begin{equation}
	\lim_{\gamma \to \infty}
	\varepsilon^{\mathrm{th}}(\frac{d}{d_{\max}})
	=
	\varepsilon_{\mathrm{pre}}^{\mathrm{th}}, \quad d>0,
	\label{eq:thermal_inf}
\end{equation}
while $\varepsilon^{\mathrm{th}}(0)=0$ for all $\gamma$.
This limit corresponds to a strongly localized thermal strain near the kirigami pattern boundary.

\section{Appendix B. Equivalent Mooney-Rivlin Parameters from Infinitesimal Strain Linearization}
\label{app:MR}

For linear elasticity, we have
$\sigma_{ij} = C_{ijkl}\,\varepsilon_{kl}$. 
For an isotropic material, this relation can be written in Lam\'e form as
\begin{equation*}
	\boldsymbol{\sigma}
	=2\mu\,\boldsymbol{\varepsilon}
	+ \lambda(\operatorname{tr}\boldsymbol{\varepsilon})\mathbf I,
\end{equation*}
where $\mu$ and $\lambda$ are the Lam\'e parameters. Equivalently, the Cauchy stress can be expressed as
\begin{equation}
	\boldsymbol{\sigma}
	= 2\mu\,\boldsymbol{\varepsilon}'
	+ \left(\lambda + \frac{2\mu}{3}\right)
	(\operatorname{tr}\boldsymbol{\varepsilon})\mathbf I
    = 2G\,\boldsymbol{\varepsilon}'
	+ K(\operatorname{tr}\boldsymbol{\varepsilon})\mathbf I,
    \label{eq:Hookes_law}
\end{equation}
where, the deviatoric strain $\boldsymbol{\varepsilon}'
= \boldsymbol{\varepsilon}
- \frac{1}{3}(\operatorname{tr}\boldsymbol{\varepsilon})\mathbf I$, 
shear modulus $G=\mu$,
and bulk modulus $K=\lambda + \frac{2}{3}\mu$.

\medskip
With Equation~\ref{eq:Hookes_law}, the linear elastic energy density is given by
\begin{equation}
	W_{\mathrm{lin}}
	= \frac12\,\boldsymbol{\sigma}:\boldsymbol{\varepsilon}
	= \frac12
	\Big(
	2G\,\boldsymbol{\varepsilon}'
	+ K(\operatorname{tr}\boldsymbol{\varepsilon})\mathbf I
	\Big)
	:\boldsymbol{\varepsilon}
	= G\,\boldsymbol{\varepsilon}':\boldsymbol{\varepsilon}'
	+ \frac{K}{2}(\operatorname{tr}\boldsymbol{\varepsilon})^2,
    \label{eq:energy_linear}
\end{equation}
where Equation~\ref{eq:Hookes_law} was substituted, and 
\(
\boldsymbol{\varepsilon}':\mathbf I = 0
\)
and
\(
\mathbf I:\boldsymbol{\varepsilon}=\operatorname{tr}\boldsymbol{\varepsilon}
\) were used.

\medskip
For an isotropic, compressible hyperelastic material, the Mooney-Rivlin strain energy density is given by \cite{dassault_systemes_simulia_corp_abaquscae_2020}
\begin{equation}
	W(\mathbf F)
	=
	C_1(\bar I_1-3)
	+
	C_2(\bar I_2-3)
	+
	\frac{1}{D_1}(J-1)^2,
	\label{eq:MR}
\end{equation}
where
\begin{equation*}
	\bar I_1 = J^{-2/3} I_1, \quad
	\bar I_2 = J^{-4/3} I_2, \quad
    I_1=\mathrm{tr}\,\mathbf C, \quad
	I_2=\tfrac12[(\mathrm{tr}\,\mathbf C)^2-\mathrm{tr}(\mathbf C^2)], \quad
    J = \det \mathbf F, \quad
    \mathbf C=\mathbf F^{T}\mathbf F.
\end{equation*}

\medskip
The infinitesimal strain limit of Equation~\ref{eq:MR} gives 
\cite{abeyaratne_continuum_1998, holzapfel_nonlinear_2002}
\begin{equation}
	W
	=
	2(C_1 + C_2)\,\varepsilon':\varepsilon'
	+
	\frac{1}{D_1}\left(\mathrm{tr}\,\varepsilon\right)^2.
	\label{eq:W_MR_linear}
\end{equation}
Comparing Equations~\ref{eq:energy_linear} and \ref{eq:W_MR_linear}, the equivalent shear and bulk moduli are given by
\begin{equation}
	G=2(C_1+C_2), \qquad K=2/D_1.
\end{equation}
Assuming $C_1=C_2$, we obtain $C_1=C_2=G/4$.

}

\section{Appendix C. Dimensions of Bilayer Composites}
\label{app:dimensions}

\begin{figure}[!htb]
\begin{center}
{\includegraphics[width=6.5in]{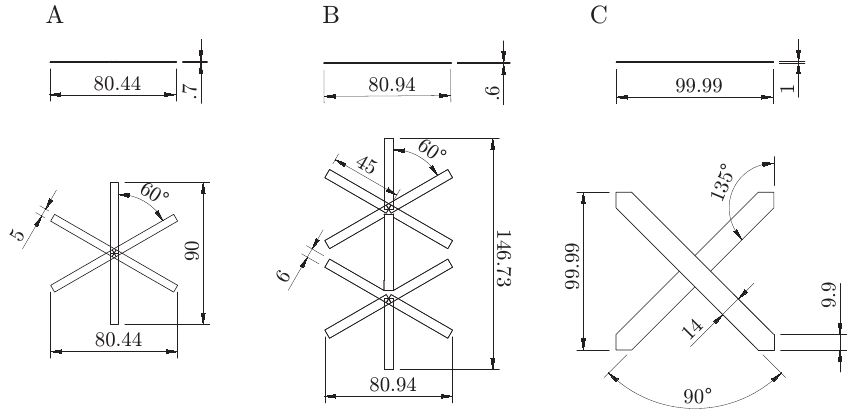}}
\caption{Dimensions of the three representative bilayer composites in mm.}
\label{fig:10}
\end{center}
\end{figure}

The dimensions of the three representative bilayer composites
are shown in Figure~\ref{fig:10}.

\medskip

\bibliographystyle{MSP}
\bibliography{AMT25}

\end{document}